\documentclass[jkps,fleqn,showpacs,showkeys]{revtex4}
\usepackage{graphicx}
\usepackage{amssymb}
\usepackage{amsmath}
\usepackage{bm}
\begin{document}
\setcounter{page}{0}
\title[]{Numerical Study on Quantum Walks Implemented on the Cascade Rotational Transitions in a Diatomic Molecule}
\author{Leo \surname{Matsuoka}}
\email{matsuoka.leo@jaea.go.jp}
\thanks{Fax: +81-774-71-3338}
\author{Tatsuya \surname{Kasajima}}
\author{Masashi \surname{Hashimoto}}
\author{Keiichi \surname{Yokoyama}}
\affiliation{Quantum Beam Science Directorate, Japan Atomic Energy Agency, Kyoto 619-0215, Japan}

\date[]{}

\begin{abstract}
We propose an implementation scheme for the continuous-time quantum walk using a diatomic molecule and an optical frequency comb.
We show an analogy between the quantum walk and the cascade rotational transitions induced by the optical frequency comb
whose frequency peaks are tuned to the pure rotational transitions in the molecule.
The strategy to compensate for the centrifugal distortion of the real molecule is also demonstrated.
\end{abstract}



\maketitle

Submitted to: J. Kor. Phys. Soc., Proceedings of AISAMP9

\section{INTRODUCTION}
The quantum walk is the quantum counterpart of the well-known classical random walk.
The quantum walk has been actively studied for the quantum information processing\cite{kempe_review,konno_note}.
Historically, two types of quantum walks have been studied independently,
that is, the continuous-time quantum walk (CTQW)\cite{farhi_CTQW_pra} and the discrete-time quantum walk (DTQW)\cite{aharonov_DTQW_pra}.
The CTQW describes a continuous diffusion of the probability amplitude on a series of discrete sites,
whereas the DTQW describes an expansion of the entanglement by the discrete unitary operations on a series of qubits. 
In these days, the concept of the CTQW attracts more and more attention.
For example, the CTQW can be regarded as a universal computational primitive\cite{childs_univ_prl}.
Also, the concept of the CTQW helps us to understand deeply the energy transfer mechanism in photosynthesis\cite{engel_photosynthesis_nature,mohseni_photosynthesis_jcp}.

The implementation schemes of the CTQW were studied by some groups.
The 4-state cyclic CTQW was realized first on a two-qubit nuclear-magnetic-resonance quantum computer\cite{du_NMRCTQW_pra}.
Schemes based on the quantum dots\cite{solenov_qdot_pra}, and Rydberg atoms in an optical lattices\cite{cote_rydberg_newphys} were also suggested.
Recently, the large scale one-dimensional CTQW was realized by using waveguide lattices\cite{perets_waveguide_prl}.
The waveguide lattices can be used for the advanced information processing using correlated photons\cite{peruzzo_correlate_sci}.
The implementation of the quantum walk should facilitate the experimental simulation of various phenomena,
even if that cannot be used for the correlated particles.
To be used for the simulation system, both the scalability and the controllability are required for the implementation scheme.
To determine the probability distribution of the CTQW of many steps,
parallel execution of a huge number of trials is favorable
because the number of required trial rapidly increases as the number of steps increases.
Also, to implement various modifications, the driving force of the CTQW should be controllable.
In this paper, we propose a new implementation scheme of the CTQW using a diatomic molecule and an optical frequency comb.
Both the controllability and the scalability are expected to be included in this scheme.

\section{THE CONTINUOUS-TIME QUANTUM WALK }
We briefly review the definition of the CTQW.
In one dimensional space, the CTQW is expressed as the finite-difference Schr\"odinger equation
\begin{equation}
\label{CTQW_define}
i\frac{\partial}{\partial t}\psi(n,t)=\gamma\left [-\frac{1}{2}\psi(n-1,t)-\frac{1}{2}\psi(n+1,t)\right ],
\end{equation}
where $\psi(n,t)$ is a complex amplitude at the continuous time $t$
and discrete site position $n$, and $\gamma$ is the rate constant
of the transition between neighboring sites.
With the initial condition of $\psi(n,0)=\delta_{n,0}$, the exact solution of CTQW is given by
\begin{equation}
\label{CTQW_bessel_w}
\psi(n,t)=i^{|n|}\mathcal{J}_{|n|}(\gamma t),
\end{equation}
\begin{equation}
\label{CTQW_bessel}
|\psi(n,t)|^2=\mathcal{J}_{|n|}^2(\gamma t), 
\end{equation}
where $\cal{J}$$_n$ is the $n$-th order Bessel function of the first kind\cite{konno_note}.
The probability distribution of the CTQW evolves quite differently from its classical counterpart (Fig.1).
The main features of the CTQW, compared with the classical random walk,
are the peaks that stands at both ends,
the rapid diffusion of the probability distribution, and the sharp reflecting behavior at the boundary.

\begin{figure}
\includegraphics[width=7.5cm,keepaspectratio]{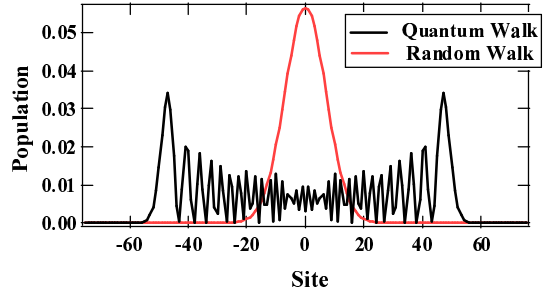}
\caption{(Color online) An example of probability distributions of the continuous-time classical random walk and the CTQW.
}\label{fig1}
\end{figure}

\section{PHYSICAL IMPLEMENTATION USING DIATOMIC MOLECULES}
\label{imple}
Diatomic molecule is one of the most characterized multi-level system.
First, we regard the rotational states of a diatomic molecule as the nodes of the quantum walk.
Due to the selection rule of $\Delta J =\pm 1$, the rotational states compose a one-dimensional network
which is connected by the pure rotational transitions.
When the molecules are irradiated by the electromagnetic wave which can induce both $\Delta J =\pm 1$ transitions,
the molecules are excited, de-excited, or remained as it is according to a certain probability.
Furthermore, if the electromagnetic wave induces all the rotational transitions in the network,
the irradiated molecules are repeatedly excited or de-excited, as if they were randomly walking on the network.
However, the evolution dynamics of them is not the random walk but the quantum walk.
To induce all the rotational transitions in the molecule, a broad-band laser pulse seems to be suitable.
Actually, however, the broad-band laser pulse usually induces multi-photon processes such as ionization.

To avoid the undesired processes,
we consider the optical frequency comb which can induce only a series of the pure rotational transitions (Fig.2).
We analytically show the analogy between the CTQW and the cascade pure rotational transitions
induced by the optical frequency comb in the molecules.
At first, we discuss an ideal situation, in which the centrifugal distortion can be fully ignored.
The rotational energy of diatomic molecules with a closed shell structure
is expressed as $E_J=hBJ(J+1)$, where $B$ is the rotational constant.
Because the transition frequency for $J \rightarrow J+1$ is expressed as $\nu_{J}=2B(J+1)$,
such transitions exhibit an evenly spaced spectrum like a comb in the frequency domain.
Under the dipole approximation for linearly polarized electric field $\varepsilon(t)$,
the equation of rotation in the interaction picture is written by
\begin{eqnarray}
i\hbar \frac{d}{dt}c_J(t)=& -\mu_{J-1}\varepsilon (t)\exp(2\pi i \nu_{J-1}t)c_{J-1}(t) \nonumber \\
                     & -\mu_J\varepsilon (t)\exp(-2\pi i \nu_{J}t)c_{J+1}(t),
\label{sch_expand}
\end{eqnarray}
where $c_J(t)$ is the complex amplitude of the rotational state $J$,
and $\mu_J$ is the transition dipole moment for the pure rotational transition from $J$ to $J+1$,
which is calculated by $\mu_J=\mu \{\left[(J+1)^2-M^2\right]/\left[(2J+1)(2J+3)\right]\}^{1/2}$
using the permanent dipole moment $\mu$ and the magnetic quantum number $M$.
In this paper, $\mu=1.0$, $M=0$, and $\hbar=1.0$ are assumed for simplicity.

\begin{figure}
\includegraphics[width=12.0cm,keepaspectratio]{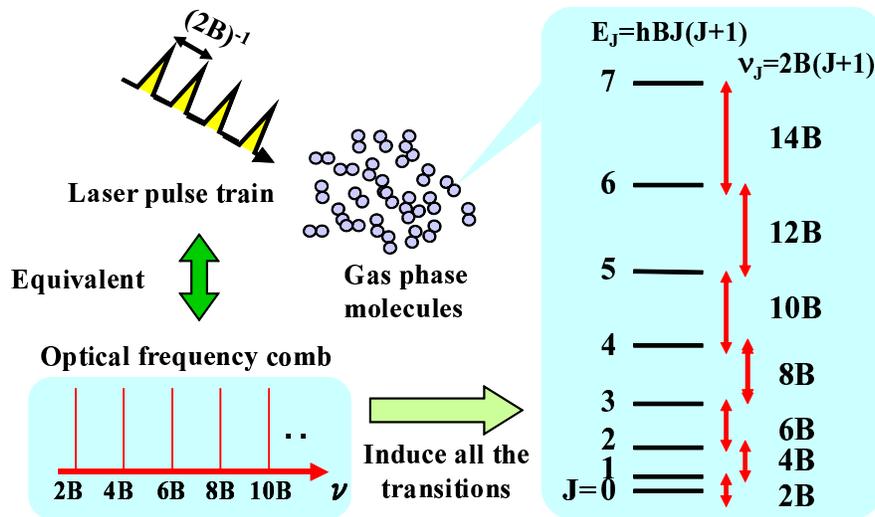}
\caption{(Color online) Conceptual picture and schematic diagram of the implementation scheme of the CTQW using the diatomic molecules
and the optical frequency comb in the ideal situation.
}\label{fig2}
\end{figure}

For the optical frequency comb, we define the electric field as
\begin{equation}
\label{field_exp}
¡¡\varepsilon (t) =  \sum_{J'=0}^{J_{max}-1} \frac{\gamma}{\mu_{J'}}\cos(2\pi \nu_{J'}t).
\end{equation}
The $\varepsilon (t)$ is regarded as a train of pulses whose interval is $(2B)^{-1}$ in the time domain (Fig.3).

\begin{figure}
\includegraphics[width=7.5cm,keepaspectratio]{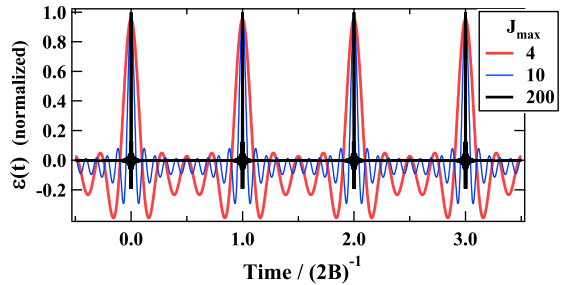}
\caption{(Color online) Time profiles of the optical frequency comb with various numbers of frequency component $J_{max}$.
}\label{fig3}
\end{figure}

Using eq.\ref{field_exp} and
putting all the oscillating terms into term A (rotating wave approximation), eq.\ref{sch_expand} is re-written by
\begin{equation}
\label{sch_nonres}
i \frac{d}{dt}c_J(t) = -\frac{\gamma}{2}c_{J-1}(t) -\frac{\gamma}{2}c_{J+1}(t) - A.
\end{equation}
The first two terms in eq.\ref{sch_nonres} represent the contribution from the resonant frequencies (resonant contribution),
whereas the third term A represents the contribution from the detuned frequencies (detuning contribution).
The resonant contribution has the same form as the definition of the CTQW, eq.\ref{CTQW_define},
by letting $c_J(t)=\psi(n,t)$ and $\Delta J=n$.

The detuning contributions are expected to be averaged out to zero by integrating over the duration of $(2B)^{-1}$
because all the exponential terms in A seems to oscillate with a period of $(2B)^{-1}$.
We confirmed this speculation numerically.
Eq.\ref{sch_expand} and eq.\ref{field_exp} were solved by the fourth order Runge Kutta algorithm
with the initial condition of $c_J(-0.5/2B)=\delta_{J,100}$, $J_{max} = 200$, and $\gamma =2B$.
We have found that the simulated population distribution after the 50-th pulse
completely overlapped with the exact solution as shown in Fig.4.
Therefore, we conclude that the optical frequency comb implements the CTQW on the rotational states
in a diatomic molecule under the ideal situation.

\begin{figure}
\includegraphics[width=7.5cm,keepaspectratio]{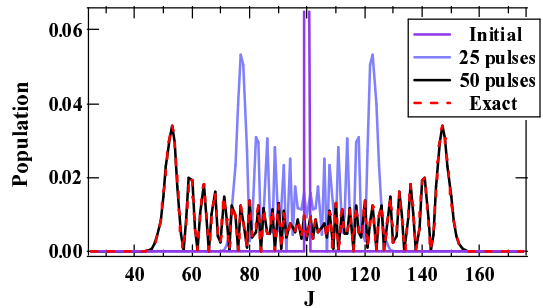}
\caption{(Color online) Simulation of the evolution induced by the optical frequency comb in the ideal situation.
Exact solution in $t=50$ is completely overlaps to the calculated lines.
}\label{fig4}
\end{figure}

Also, we briefly discuss the contribution of the two-photon transition, which has been ignored in the present calculation.
For the case of CsI molecule, which we plan to take as a test molecule in the first experimental demonstration,
the electric field of a single pulse needs to be $\sim 90$ $kV/cm$, corresponding $\sim 4.3 \times 10^{7}$ $W/cm^{2}$ in the peak intensity.
Under the supposed electric-field, the interaction potential of the two-photon transition is
expected to be smaller than that of the one-photon transition by a factor of $\sim 10^{4}$\cite{friedrich_combine_jpca}.
The probability of the net two-photon transitions can be estimated
from the product of the interaction potential and the number of the possible combination of the frequency components
which satisfy the selection rule of $\Delta J =\pm 2$.
In this case, the number of possible combination does not exceed $J_{max}$$(\simeq 10^2)$.
The net probability of the two-photon transitions is still smaller by a factor of $\sim 10^2$ than that of the one-photon transition.
We conclude that inclusion of the two-photon transition does not substantially change the probability distribution.

\section{COMPENSATION FOR THE CENTRIFUGAL DISTORTION}
For the case of real molecules,
the pure rotational transition frequencies do not exactly compose the evenly-spaced comb
because of the centrifugal distortion.
To compensate for the difference, we have to modify the electric field defined in eq.\ref{field_exp}.
The rotational energy of the real diatomic molecules is expressed as
$E'_J=hBJ(J+1)-hDJ^2(J+1)^2$, where $D$ is the centrifugal distortion constant.
The transition frequency for $J \rightarrow J+1$ is expressed as $\nu'_{J}=2B(J+1)-4D(J+1)^3$.
(For the case of CsI molecule, the ratio $D/B = 1.57 \times 10^{-7}$\cite{rusk_milli_pr}.)
The equation of rotation is re-written by
\begin{eqnarray}
i \frac{d}{dt}c_J(t)=& -\mu_{J-1}\varepsilon (t)\exp(2\pi i \nu'_{J-1}t)c_{J-1}(t) \nonumber \\
                     & -\mu_J\varepsilon (t)\exp(-2\pi i \nu'_{J}t)c_{J+1}(t).
\label{sch_cent}
\end{eqnarray}
In a similar way, the electric field for the CTQW is defined as
\begin{equation}
\label{field_cent}
¡¡\varepsilon (t) =  \sum_{J'=0}^{J_{max}-1} \frac{\gamma}{\mu_{J'}}\cos(2\pi \nu'_{J'}t).
\end{equation}
The $\varepsilon (t)$ is regarded as a train of {\it chirped} pulses
whose chirp rate gradually changes from negative to positive (Fig.5).
Eq.\ref{sch_cent} and eq.\ref{field_cent} were numerically solved by the same way as in Sec.\ref{imple} with 
the initial condition of $c_J(-12.5/2B)=\delta_{J,100}$, and $D/B = 1.57 \times 10^{-7}$.
The calculated population distribution and the exact solution were most agreed with each other (Fig.6).
We conclude that the modified optical frequency comb still approximately implements the CTQW on the rotational states
in a diatomic molecule including the effect of the centrifugal distortion.
Controlling the chirp rate in the visible region is not difficult,
and that in the mid-infrared region has been performed\cite{tsubouchi_infrared_opcom}.
That will be also possible in the terahertz region in the near future.

\begin{figure}
\includegraphics[width=12.0cm,keepaspectratio]{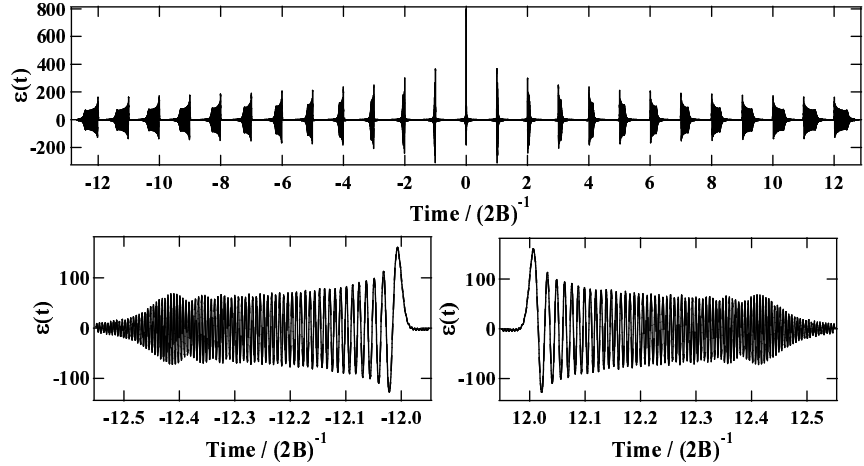}
\caption{Upper panel: Time profile of the modified optical frequency comb ($J_{max}=200$, $D/B = 1.57 \times 10^{-7}$, 25 pulses.)
Lower left panel: An expanded time profile of the first pulse.
Lower right panel: An expanded time profile of the last pulse.
}\label{fig5}
\end{figure}

\begin{figure}
\includegraphics[width=7.5cm,keepaspectratio]{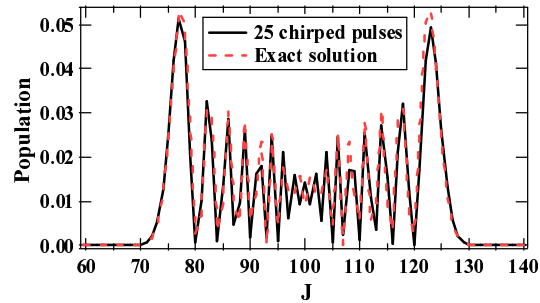}
\caption{(Color online) Simulation of the population distribution irradiated
by the modified optical frequency comb including the compensation of the centrifugal distortion ($D/B = 1.57 \times 10^{-7}$).
}\label{fig6}
\end{figure}

\section{CONCLUSIONS} 
We introduced a proposal for the physical implementation of the continuous-time quantum walk
using diatomic molecules and the optical frequency comb.
We demonstrate the analogy between the CTQW and the cascade rotational transitions in the diatomic molecules.
Possible imperfection by the centrifugal distortion could be almost eliminated by chirping each pulse appropriately.
The implementation of the CTQW in a molecule will be useful for simulating the diffusion phenomena in quantum system.

\begin{acknowledgments}
We would like to acknowledge useful discussion with Akira Ichihara.
\end{acknowledgments}

\end{document}